\documentstyle[multicol, tabularx, epsfig, amssymb]{article}
\begin{document}
\title{\textbf{Final remarks to our study of $\eta$-photoproduction on protons in the
resonance region}}
\author{V.A. Tryasuchev\footnote{Electronic address: \textbf{trs@npi.tpu.ru}}}
\date {\textsl{ Tomsk Polytechnic University, Tomsk, Russia}}
\maketitle

\begin{abstract}
The properties of previously discovered nucleon resonances are
amended basing on the recent and more detailed experimental data
about photoproduction of $\eta$-mesons on protons.
\end{abstract}


\begin{multicols}{2}
As the experimental data on $\eta$-photoproduction on nucleons have
been accumulated [1$-$5]
\begin{equation}
    \gamma + p \rightarrow \eta + p,
\end{equation}
they were gradually analyzed within dynamical models developed in
refs. [6$-$10]. After the GRAAL measurements of beam asymmetry and
differential cross sections in the energy region 1100$-$1500 MeV
were reported, necessity of essential revision of the resonance
properties collected in [7,10] has become evident.

In order to explain all the available data for the reaction (1), new
$S_{11}$ resonance with quite definite position on the complex
energy plane was required in addition to the already known baryons
$S_{11}(1535)$ and  $S_{11}(1650)$. Information on three
$S_{11}$-resonances obtained in our works is presented in Table 1.
\begin{table*}
 \centering \caption{Parameters of $S_{11}$-resonances, extracted
from our analyses of $\gamma$$p\rightarrow$$\eta$$p$.}
\begin{tabular*}{12cm}
{|c|@{\extracolsep{\fill}}c|@{\extracolsep{\fill}}c|@{\extracolsep{\fill}}c|@{\extracolsep{\fill}}c|@{\extracolsep{\fill}}c|}\hline
$N^{\ast}$ & $W_{r}$, MeV & $\Gamma_{r}$, MeV &
$\Gamma_{p\eta}$$/$$\Gamma_{r}$ & $\beta$ & $A_{1/2}$,
GeV$^{-1/2}$\\\hline $S_{11}(1535)$ & 1535 & 156 & 0.5 & 1  &
0.110\\\hline $S_{11}(1650)$ & 1642 & 140 & 0.1 & $-$1 &
0.102\\\hline $S_{11}(1830)$ & 1828 & 150 & 0.1 & 1  & 0.032\\\hline
\end{tabular*}
\end{table*}
Here, $\beta$ stands for the sign of the ratio of $\eta$\textit{NR}
and $\pi$\textit{NR} couplings. The properties of all resonances,
needed to describe the measured observables of the process (1), are
listed in Table 2. There, the absolute values $\xi_{\lambda}$ [11]
are proportional to the contributions of the corresponding
resonances to the reaction (1). As one can see, the contribution of
$P_{13}(1720)$ appears to be very important. This resonance strongly
influences the shape of angular distributions as well as beam
asymmetry in a wide energy region up to 1.9 GeV. By now, opinions
differ widely on the role of the baryon $P_{13}(1720)$ in
$\eta$-photoproduction. For example, in refs. [12$-$13] this
resonance is shown to be insignificant, whereas in more  recent
works of  [14$-$16] quite essential contribution of $P_{13}(1720)$
has been reported. Here we would like to mention that important role
of this resonance in the reaction (1) has also been pointed out in
our previous works (see, e.g., [7]).
%
\begin{table*}
\centering \caption{Resonances needed to describe the available data
for the reaction (1). The parameter $\xi_{\lambda}$ is determined as
$\xi_{\lambda}=
\sqrt{\frac{k_{\gamma}m\Gamma_{p\eta}}{q_{\gamma}W_{\gamma}\Gamma^{2}_{\gamma}}}A_{\lambda}$.}
\begin{tabular*}{17cm}[c]{|c|@{\extracolsep{\fill}}c|@{\extracolsep{\fill}}c|@{\extracolsep{\fill}}c|@{\extracolsep{\fill}}c|@{\extracolsep{\fill}}c|@{\extracolsep{\fill}}c|}\hline
$N^{\ast}$ & $W_{r}$, MeV & $\Gamma_{r}$, MeV & $\gamma^{E}$, MeV &
$\gamma^{M}$, MeV & $\xi^{1/2}$, $10^{-1}$ GeV$^{-1}$ & $\xi^{3/2}$,
$10^{-1}$ GeV$^{-1}$\\\hline $S_{11}(1535)$ & 1535 & 156 & 2.150   &
$-$      & 2.476  & $-$      \\\hline $S_{11}(1650)$ & 1642 & 140 &
$-$0.652  & $-$       & 0.837  & $-$      \\\hline $S_{11}(1830)$ &
1828 & 150 & 0.180   & $-$       & 0.216  & $-$      \\\hline
$P_{11}(1440)$ & 1440 & 350 & $-$       & 0.250   & $-$      & $-$
\\\hline $P_{11}(1710)$ & 1710 & 100 & $-$       & 0.020   & 0.022  &
$-$
\\\hline
$P_{13}(1720)$ & 1730 & 185 & $-$0.085  & 0.560   & 0.245  & 0.630
\\\hline
$D_{13}(1520)$ & 1520 & 120 & 0.300   & 0.300   & $-$0.017 & 0.145
\\\hline
$D_{15}(1675)$ & 1675 & 110 & 0.085   & $-$0.002  & 0.100  & 0.053
\\\hline
$F_{15}(1680)$ & 1685 & 130 & 0.195   & 0.075   & 0.146  & 0.131
\\\hline
$F_{17}(1990)$ & 1980 & 290 & 0.010   & 0.255   & 0.107  & 0.128
\\\hline
$G_{17}(2190)$ & 2240 & 425 & $-$0.480  & $-$0.001  & $-$0.148 &
$-$0.192
\\\hline
$H_{19}(2220)$ & 2240 & 425 & $-$0.730  & $-$0.001  & $-$0.155 & $-$0.191 \\
\hline
\end{tabular*}
\end{table*}
The three resonances
\begin{equation}
    P_{11}(1710),D_{15}(1675),F_{15}(1680)
\end{equation}
are shown to be less significant, but their inclusion results in
better description of the structural details of the observed cross
section.

As for the heavier resonances
\begin{equation}
    F_{17}(1990),G_{17}(2190),G_{19}(2250),H_{19}(2220),
\end{equation}
among which the last three are marked with four stars in the PDG
compilation [17], we found that $G_{19}(2250)$ provides only small
fraction of the resulting cross section. The other resonances, being
much more important, govern the shape of $\eta$ angular
distributions at photon energies \textit{K}$_{0}$ $>$ 1.5 GeV. At
the same time, the masses and widths of these states are not
uniquely determined. Polarization measurements in the appropriate
energy region are necessary for a more precise determination of
these parameters.

Several results coming out from our analysis seem to be unrealistic
and need further clarification. Firstly, the $S_{11}(1650)$
photoexcitation amplitude has too large magnitude which appears to
be comparable to that of $S_{11}(1535)$ (see Table 1). Secondly, our
model favors equal values of the electromagnetic amplitudes for
$F_{15}(1680)$
\begin{center}
$A_{1/2} = A_{3/2}$.
\end{center}

\begin{figure*}[ht]
   \centering
      \includegraphics[width=12.4cm,height=23cm]{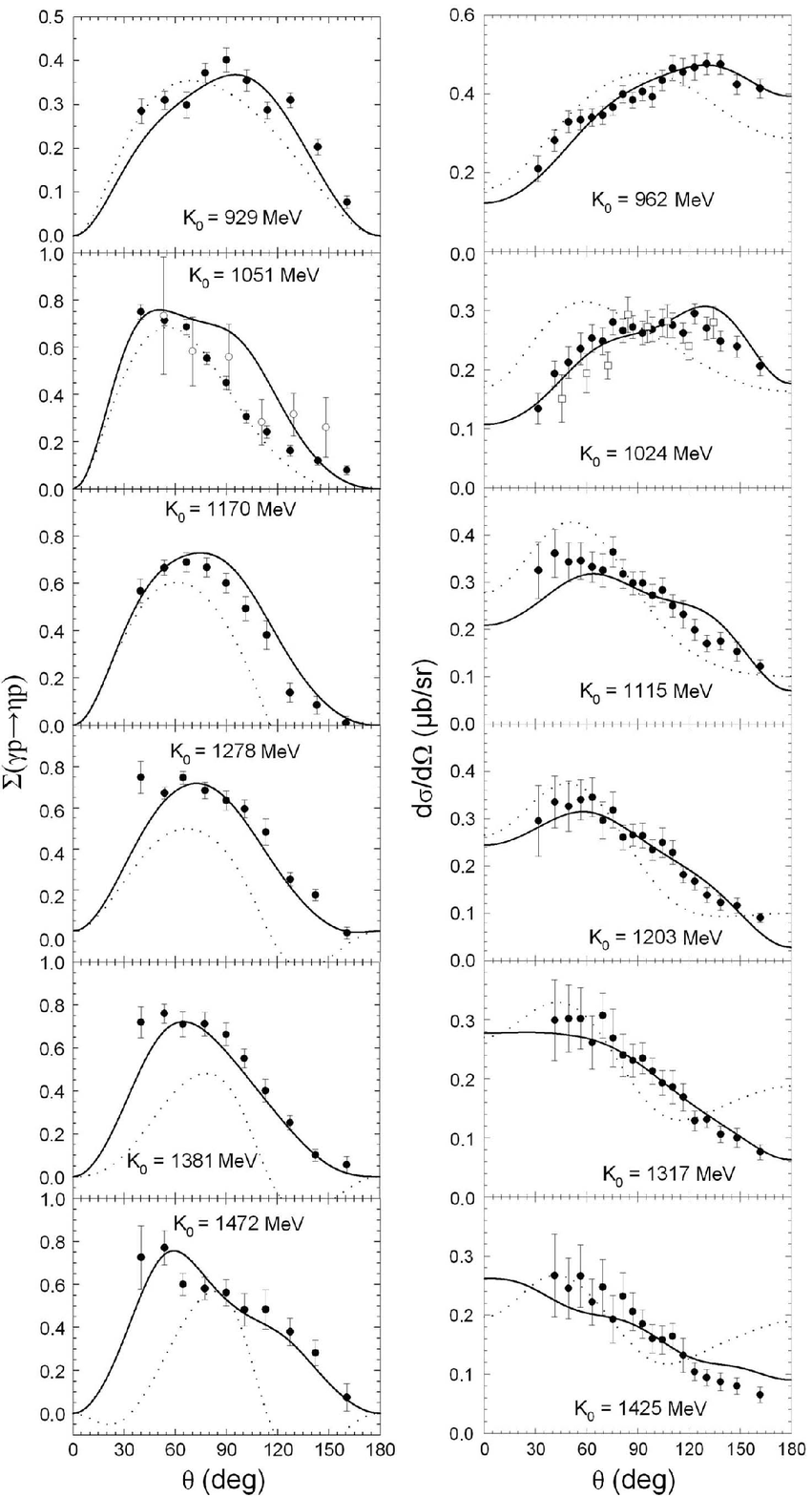}
    \caption{Beam asymmetry $\Sigma$ and differential cross section
d$\sigma$/d$\Omega$ for $\gamma$$p$$\rightarrow$$\eta$$p$. The solid
and the dotted curves are obtained in the present analyses and in
ref. [9] respectively. The data are taken from the following
references: \small $\square$ \normalsize , ref. $[2]$; \Large
$\circ$ \normalsize , ref. [4]; \Large $\bullet$ \normalsize , ref.
$[5]$. The data of ref. [4] are shown with total error.}
    \label{fig:fig1}
\end{figure*}

This paper is in contradiction with the average PDG results [17],
where $A_{1/2} << A_{3/2}$. We did not show our calculation for
photon energies below 930 MeV, since in this region, where the
reaction is dominated by $S_{11}(1535)$, the theoretical description
of the data is always good. Furthermore, not presented are the
results for beam asymmetry from ref. [4], being in well agreement
with the GRAAL data [5], with the exception of \textit{K}$_{0}$ $=$
1050 MeV.

\begin{figure*}[ht]
  \centering
     \includegraphics[width=14cm,height=12.6cm]{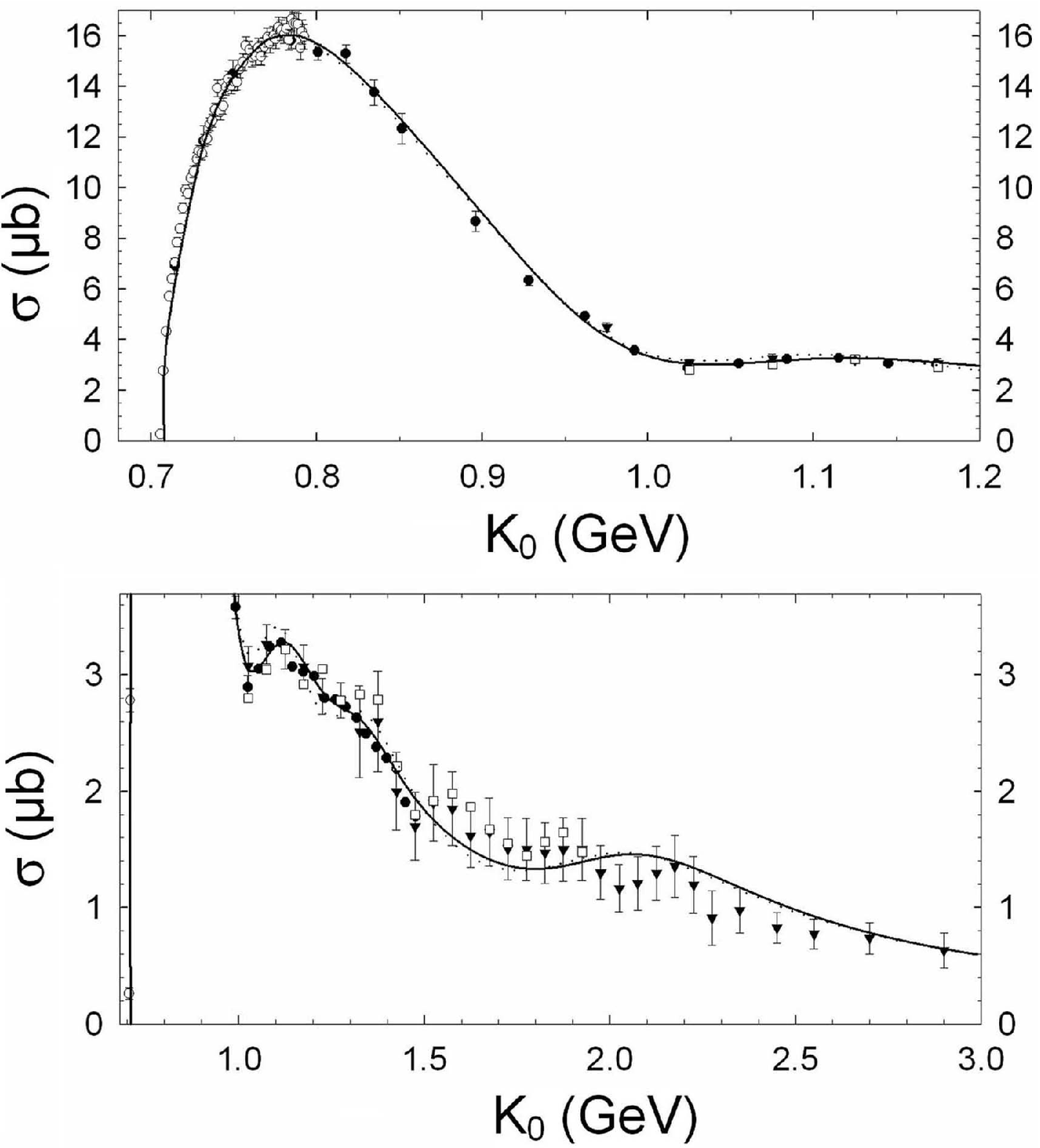}
\caption{Total cross section for the reaction
$\gamma$$p$$\rightarrow$$\eta$$p$ as function of the photon lab
energy \textit{K}$_{0}$. Notation of the curves as in Fig.1.  The
data are from: $[1]$, \Large $\circ$ \normalsize ; $[5]$, \Large
$\bullet$ \normalsize ; $[2]$, \small $\square$ \normalsize ; $[3]$,
$\blacktriangledown$.}
    \label{fig:fig2}
\end{figure*}

The region close to \textit{K}$_{0}$ $=$ 1050 MeV is just the energy
at which the measured asymmetry $\Sigma$ from ref. [4] along with
the experimental results from ref. [5] are depicted in Fig.1.

The energy dependence of the observed total cross section for the
reaction (1) is compared with all available data in Fig.2. One sees
well agreement which however is not the governing factor for
extracting the resonance parameters. The reason is presumable model
dependence of the experimental results caused by the limited range
of polar angles of particles detected in the GRAAL measurements.

We think the conclusion about total agreement between the
calculation and the data would be premature, even if only the
limited region of the photon energy ($<$ 1.5 GeV) is considered. In
Fig.3 we present beam asymmetry of the reaction (1) as function of
the photon energy at different angles of $\eta$-mesons.

\begin{figure*}[ht]
  \centering
     \includegraphics[width=12.6cm,height=10cm]{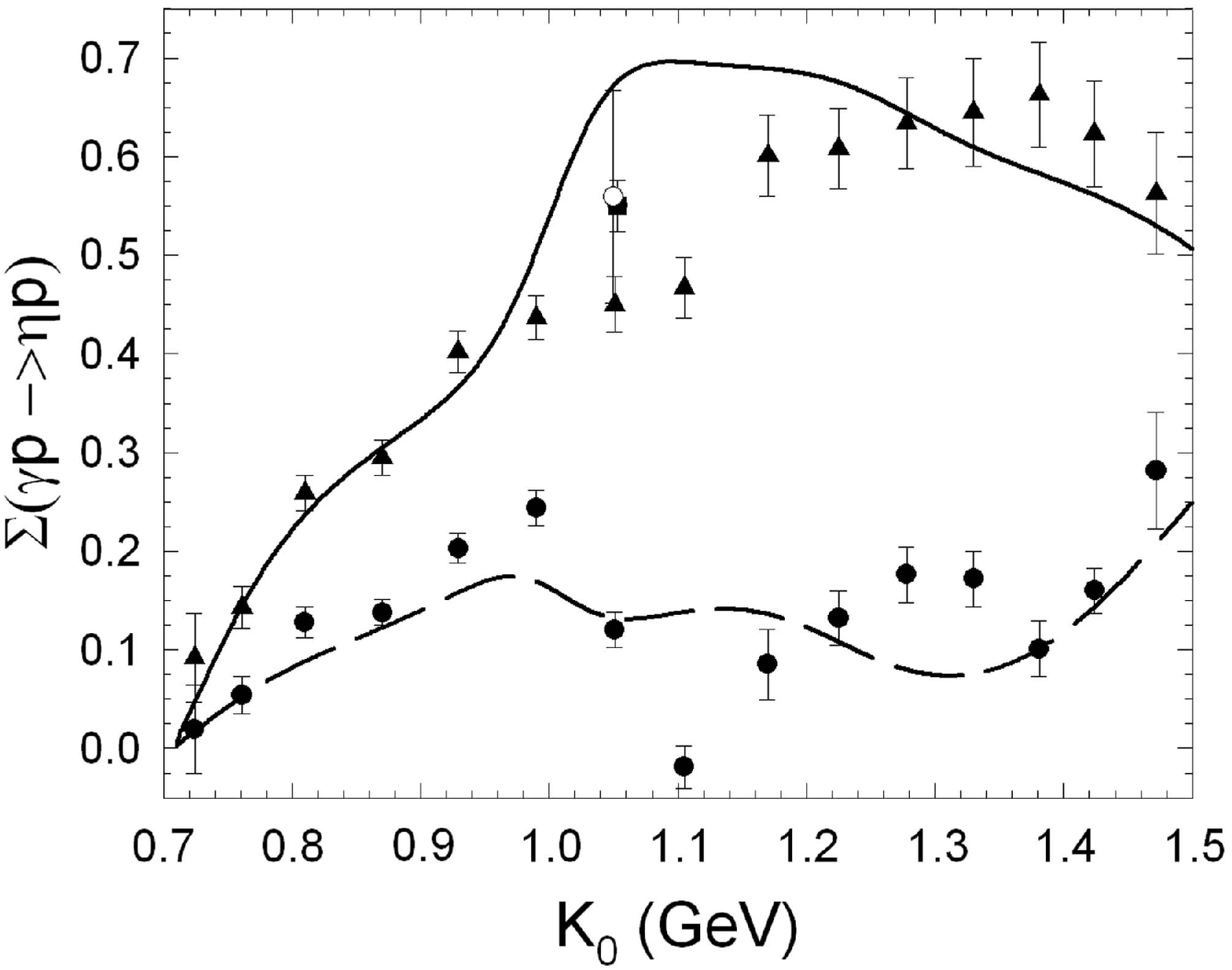}
\caption{Beam asymmetry for $\gamma$$p$$\rightarrow$$\eta$$p$ as
function of photon energy shown for two \textbf{$\eta$} c.m. angles:
90$^{\circî}$, \normalsize $\blacktriangle$ $-$ data $[5]$ (solid
curve is our calculation); 142.5$^{\circî}$, \Large $\bullet$
\normalsize $-$ data $[5]$ (our result is shown by the dashed
curve). For $\theta = 90^{\circî}$ the data are shown by: \Large
$\circ$ \normalsize , from ref. $[4]$ and \small $\blacksquare$
\normalsize , from ref. $[18]$.}
   \label{fig:fig3}
\end{figure*}

As is pointed out in [5] and may also be seen in Fig.3, in the
region 1.05$-$1.2 GeV the data exhibit anomalous behavior, in
particular, at $\theta$ = 90$^{\circî}$ and 142.5$^{\circî}$. Here
one also notes the maximum deviation between the final GRAAL data
[5] and their preliminary results [18], as well as with the results
of ref. [4]. This feature was also noticed by the authors of ref.
[19] who were searching for signatures of a narrow ($\Gamma$ $<$ 25
MeV) resonance $P_{11}$ in meson photoproduction in the second
resonance region. Since this resonance is assumed to be the member
of the \textit{SU}(3) flavor antidecuplet (\textit{J} $=$ $1/2$),
one can easily show that due to conservation of the
$\textit{U}$-spin this state is photoproduced mostly on neutrons,
rather than on protons. As is also noticed in refs. [19,20] its
existence should result in an anomalous behavior of the beam
asymmetry of the process (1) around the invariant energy $W=1680$
MeV.

Our description of the data at these energies is also inadequate.
Direct calculation shows that in the region discussed, the most
important contribution comes form $D_{15}(1675)$, $F_{15}(1680)$,
whose properties are not well known so far. This however may be
direct consequence of the above mentioned narrow resonance, which is
seen in the reaction $\gamma$$n$$\rightarrow$$\eta$$n$. In this
connection, further measurements of the $\Sigma$ asymmetry around
1.05$-$1.2 GeV with a better energy and angular resolution are of
special interest, as a tool to study these resonances on a higher
quantitative level.

\begin{figure*}[ht]
  \centering
     \includegraphics[width=13cm,height=12.7cm]{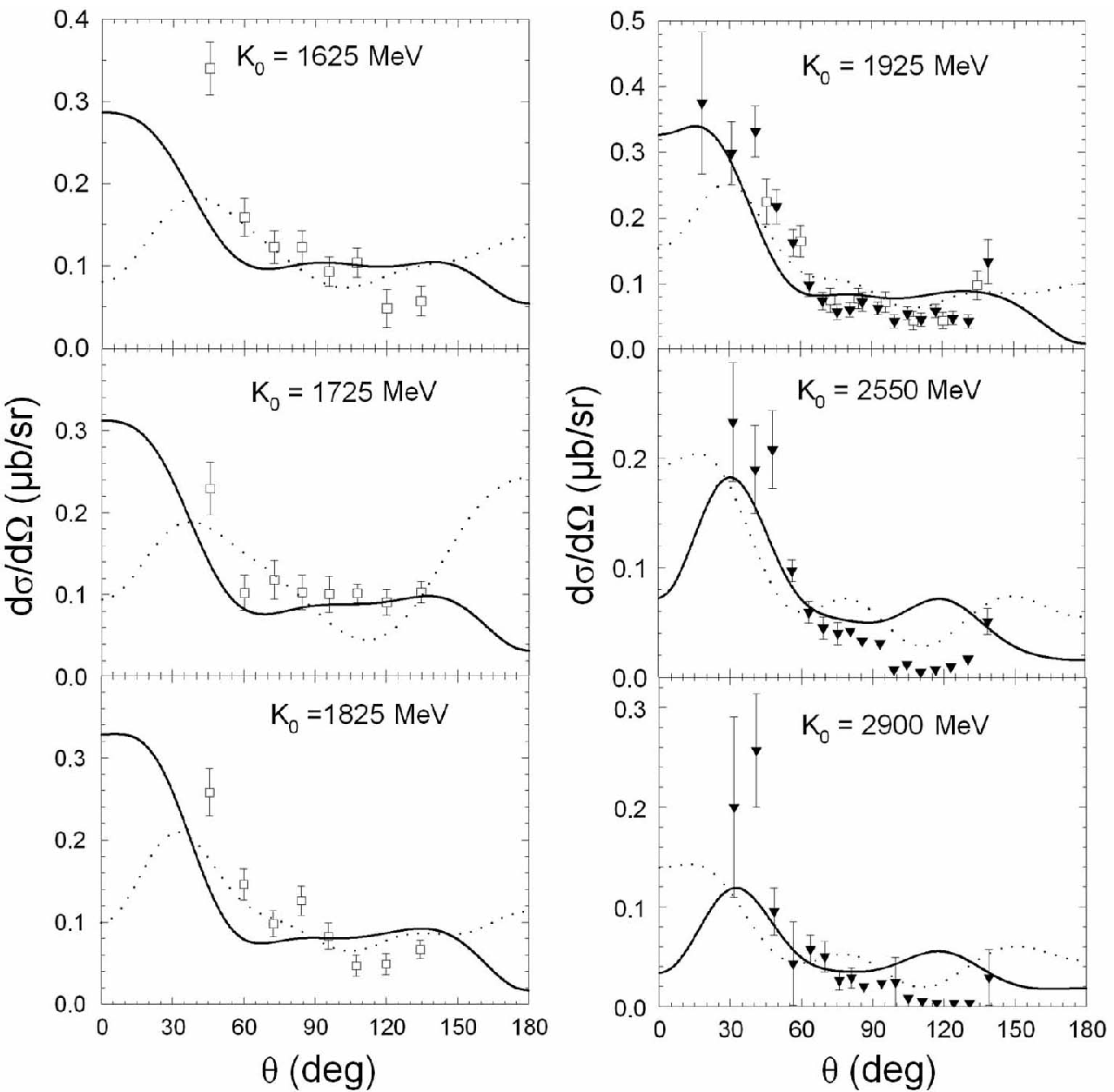}
\caption{Angular distribution of $\eta$-mesons calculated at a
photon energy above 1.5 GeV.  The meaning of the curves as in Figs.
1 and 2. Experimental data:  \small $\square$ \normalsize , from
ref. $[2]$; $\blacktriangledown$ , from ref. $[3]$.}
    \label{fig:fig4}
\end{figure*}

Above 2.2 GeV, experimental uncertainty of photon energy becomes
large, so that it is reasonable to analyze only the gross structure
of the cross section without trying to reproduce the detailes. At
the same time, in this case the observed angular dependence for
$\theta$ $<$ 30$^{\circ}$  may deserve attention as a way to
identify the energy region where the diffraction mechanism of
$\eta$-production starts to come into play. The case in point is
apparent shifting of the angular distribution to the forward
hemisphere (see Fig.4) which might bear witness to significance of
the \textit{t}-channel. In most of the isobar models, the
\textit{t}-channel mechanisms are suppressed by small coupling
constants and/or sharp formfactors. Similar to the present
theoretical base, in these models the VNN vertices and cut-offs are
usually fitted to the data for the reaction (1) at lower energies
(\textit{K}$_{0}$ $<$ 1.5 GeV). At the same time, inclusion of only
the resonances (3) turns out to be insufficient to account for the
peak in the cross-section at forward angles (see Fig.4). Additional
resonances are required to reproduce this characteristic shape.

It is worth to note a successful attempt in refs. [15,16,21] to
achieve global description of $\eta$-photoproduction on protons.
Here rather well agreement with the data is obtained primarily due
to additional inclusion of new resonance states (more than three
resonances in [15,16]). Or, straight conversely, the number of
resonances was artificially reduced in order to make the
$\chi^{2}$-method [21] more effective, so that fitted  parameters
may further be used in other reactions with $\eta$-mesons.

In conclusion, we presented our analysis of the final experimental
results obtained at GRAAL. Special emphasis is put on those aspects
of the process (1) which require additional experimental and
theoretical investigations. New data are of special interest for
further improvements of our knowledge about the known resonances as
well as those, whose existence is still under debate.

The author thanks A.I. Fix and A.V. Isaev for enlightening
discussions.

\end{multicols}
\end{document}